\documentclass[english]{article}
\usepackage[T1]{fontenc}
\usepackage[latin9]{inputenc}
\usepackage{color}
\usepackage[dvipsnames]{xcolor}
\usepackage{amsmath}
\usepackage{amssymb}
\usepackage{graphicx}
\usepackage{babel}

\newcommand{\be}{\begin{equation}}
\newcommand{\ee}{\end{equation}}
\newcommand{\ben}{\begin{eqnarray}}
\newcommand{\een}{\end{eqnarray}}

\newcommand{\ket}[1]{|#1\rangle}
\newcommand{\bra}[1]{\langle #1|}

\begin{document}
\title{A model of interacting quantum neurons with a dynamic synapse}
\author{J.J. Torres and D. Manzano\\
{\small{}Institute Carlos I for Theoretical and Computational Physics,
}\\
{\small{}University of Granada, Granada, E-18071 Spain}}
\maketitle
\begin{abstract}

Motivated by recent advances in neuroscience, in this work, we explore the emergent behaviour of
quantum systems with a dynamical biologically-inspired qubits interaction. We use a minimal model of two interacting qubits with an activity-dependent dynamic interplay as in classical dynamic synapses that induces the so-called synaptic depression, that is, synapses that present synaptic fatigue after heavy presynaptic stimulation. Our study shows that in absence of synaptic depression the 2-qubits quantum system shows typical Rabi oscillations whose frequency decreases when synaptic depression is introduced, so one can trap excitations for a large period of time. This creates a population imbalance between the qubits even though the Hamiltonian is Hermitian. This imbalance can be sustained in time by introducing a small energy shift between the qubits. In addition, we report that long time entanglement between the two qubits raises naturally in the presence of synaptic depression. Moreover, we propose and analyse a plausible experimental setup of our 2-qubits system which demonstrates that these results are robust and can be experimentally obtained in a laboratory.

\end{abstract}

\section{Introduction}

Learning and information processing are key topics of science that have recently been pushed to the quantum domain. In the last years, thanks to the development of quantum computers, there is an increasing interest in the design of autonomous devices to perform certain tasks with quantum improvement. This has lead to the developing of the fields of Quantum Machine Learning and Quantum Artificial Intelligence \cite{biamonte:n17,dunjko:rpp18}. In this direction, there have been proposal for autonomous machines that can estimate a state or a quantum unitary \cite{fischer:pra00,manzano:njp09}, as well as algorithms of quantum reinforcement learning  \cite{dong:ieee08,briegel:sr12,mautner:ngc15,yenchi:ieee20,lockwood:2020,xiao:njp22},  and quantum neuronal networks (QNN) \cite{torrontegui:epl19,chakraborty:ieee20}. Besides, the most used framework for quantum machine learning is based on Variational Quantum Algorithms (VQAs) \cite{xiao:pra20,cerezo:nwp21,mangini:epl21,bharti:rmp22}.

In the field of QNN there have been theoretical proposals of models of single quantum neurons  \cite{cao:arxiv17,tacchino:npj19,kristensen:21}, as well as networks such as the perceptron  \cite{huber:arxiv_21,pechal:arxiv21,silva:nn20,wiebe:npj16} and Hopfield's \cite{rotondo:jpa18,rebentrost:pra2018}. The main interest here has been to see if such quantum versions of neural networks are able to improve the properties of classification and pattern recognition compared with classical ones, but there is also a biological motivation \cite{adams:avs20}. All these approaches use versions of binary neurons that are substituted by qubits, with very simple interactions between the units. However, classical neural networks and  biological inspired neural population models include other important element that has been shown to have a prominent role on neural computation, i.e., the synapses \cite{amit_89}. The transmission of the information encoded in firing patterns of the neurons is performed by the synapses to postsynaptic neurons through highly non-linear processes. These include, among others, the biophysical processes that control the trafficking and recycling of neurotransmitter molecules at the synapses and which are responsible for the transmission of the electrical signals among interconnected neurons. During  the last decades, neuroscientists have extensively studied the role that synaptic processes can have on the processing of information in the brain. In particular, it has been reported in different neural media that due to the incoming presynaptic activity, synapses can reduce (synaptic depression) or increase (synaptic facilitation) their capability to transmit the incoming electrical signals \cite{tsodyks:nc98}. This clearly shows that actual synapses are \emph{dynamical} or activity-dependent entities. Moreover, such dynamic synapses have strong computational implications \cite{torres:fcn13} in the behaviour of classical neural networks, including a strong effect on storage capacity \cite{torresstorage, mejiasstorage}, the appearance of dynamical memories \cite{panticdyn,torrescompetition} and the emergence of stochastic multiresonances during the processing of weak stimuli \cite{mejiasreso}, to name a few.

With this motivation, in this work, we explore the emergent behaviour of quantum systems with a dynamical biologically-inspired qubits interaction. We use a minimal model of two interacting qubits with an activity-dependent dynamic interplay as in classical dynamic synapses. Although our study can be easily generalised for dynamic synapses which include both synaptic depression and synaptic facilitation, we here only report results concerning the case of depressing synapses, that is, synapses that present synaptic fatigue after heavy presynaptic stimulation. We observe that in absence of synaptic depression our 2-qubits quantum system shows typical Rabi oscillations. However, when synaptic depression is introduced such Rabi oscillations decrease their frequencies so one can maintain  a given qubit active for long periods of time. This creates an asymmetry between the qubits even though the Hamiltonian is Hermitian. This asymmetry can be sustained in time by introducing a small energy shift between the qubits. We also study the effect of dynamic interaction depression in the creation of entanglement between the two qubits, probing that long time entanglement raises naturally. Moreover, by analysing the behaviour of a plausible experimental setup of our 2-qubits system, we demonstrate that these results are robust and can be experimentally tested in a laboratory.

\section{Model: Two interacting qubits Hamiltonian with short-term depression}

Our model is based on two qubits with an $XY$ interaction in the form 

\begin{equation}
H(t)=\epsilon_{1}\sigma_{1}^{z}+\epsilon_2 \sigma_{2}^{z} +\frac{\Omega}{2} \, r(t) \, \left( \sigma_1^+ \sigma_2^- + \sigma_1^- \sigma_2^+ \right), 
\label{eq:ham}
\end{equation}
where $\sigma_{i}^z$ are Pauli matrices for the $i$th qubit, $\epsilon_i$ are the one-site energies, $\sigma_i^\pm$ are the creation/annihilation spin operators acting on site $i$,  $\Omega$ is a parameter characterising the qubits interaction strength, and $r(t)$ is the time-dependent parameter we will use to model the synaptic depression mechanism. By tuning the time-dependent parameter $r(t)$ the interaction between the spins can be switched on and off. This kind of $XY$ Hamiltonians have been broadly studied in different fields as quantum transport \cite{znidaric:pre11,manzano:pre12, manzano:njp16} and quantum biology \cite{caruso:jcp09,manzano:po13}.
 
\begin{figure}
\begin{centering}
\includegraphics[width=13cm]{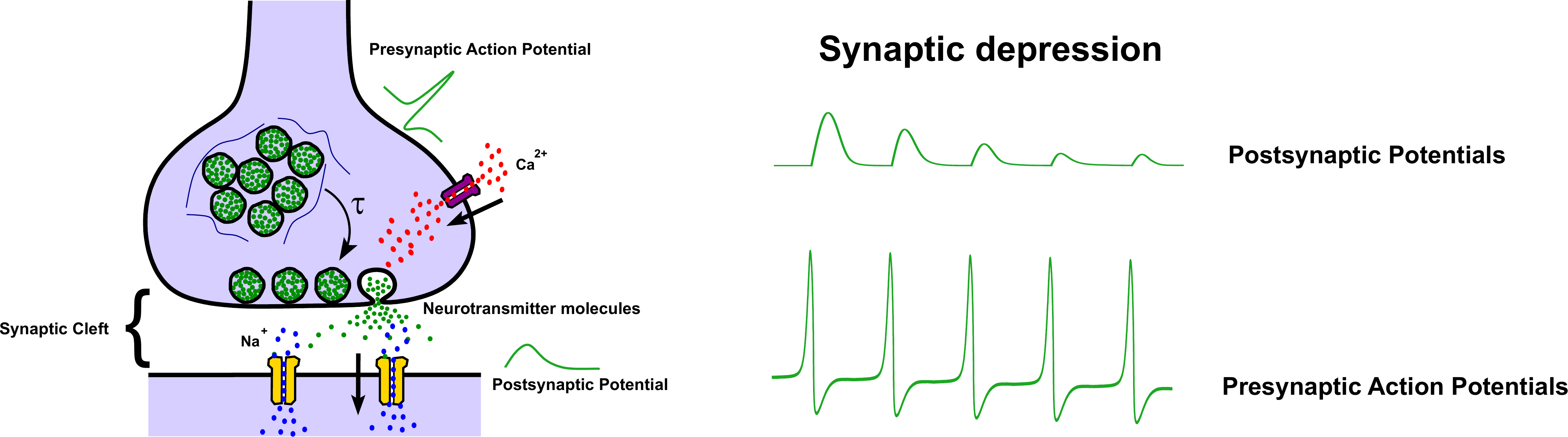}
\par\end{centering}
\caption{Sketch of the neuron-neuron interaction through an excitatory chemical synapse (left) which induces synaptic depression in the postsynaptic response (right). This occurs since the presynaptic neuron needs a time $\tau$ to move the neurotransmitter vesicles from a reserve pool of vesicles to the ready-releasable pool near the cell membrane. For large $\tau$ the number of vesicles in the ready-releasable pool are hardly replaced at the time a presynaptic action potential arrives to the synapse, which causes a decrease or depression of the postsynaptic response. This effect is stronger for large frequency of the incoming action potentials. The variable $r(t)$ in Eq. (\ref{eq:dynamic}) is related with the amount of neurotransmitters released in the synaptic cleft and determines the amplitude of the postsynaptic potentials.}
\label{figure:synapses}
\end{figure}

To tune the variable $r(t)$, we use a biologically-inspired dynamics. From the neuroscience perspective, it is now well known that the postsynaptic response of chemical synapses can vary in scales from milliseconds to minutes, in addition to more familiar long-term plastic effects due to the incoming presynaptic activity \cite{stevens:n95,markram:n96,tsodyks:pnas97,tsodyks:nc98,zucker:arp02}. Thus, synaptic efficacy can decrease due to the depletion of neurotransmitters inside the synaptic button after heavy presynaptic activity, inducing the so called \emph{short-term depression} (STD). Additionally, the postsynaptic response can be enhanced due to the growth of residual intracellular calcium concentration after the opening of the voltage gated calcium channels due to successive arrival of presynaptic action potentials to the synaptic button \cite{bertram:jn96,jackman:n17}. This effect is well known that increases the neurotransmitter release probability and the postsynaptic response, inducing the so called \emph{short-term synaptic facilitation } (STF). Both synaptic processes, i.e. STD and STF, can coexist and compete in actual synapses inducing complex emergent behaviour \cite{torrescompetition, mejiasstorage,mejiasreso,marro_21} and resulting in strong computational implications \cite{torres:fcn13}. As a first step, we are going to consider here synapses including only STD, which can be described by classically monitoring the time dependence of the fraction $r(t)$  of neurotransmitters which are ready to be released after the arrival of an action potential (in the present quantum model this fraction will modulate the interaction between both qubits, that is why it is named as the time-dependent variable of the Hamiltonian), and which follows the dynamics 
\begin{equation}
\frac{dr(t)}{dt}=\frac{1-r(t)}{\tau}-Ur(t)\delta(t-t_{sp}),
\label{eq:dynamic}
\end{equation}
where the parameter $U$ is the release probability, $\tau$ is the neurotransmitter recovering time, and $t_{sp}$ is the time at which the presynaptic spike arrives at the synapse (note that this is defined for classical systems). The presence of the Dirac delta function $\delta(t)$ in the second right-hand term of (\ref{eq:dynamic}) indicates that this term is only present at $t=t_{sp}.$ The dynamics (\ref{eq:dynamic}) implies that each time a presynaptic spike occurs, a constant portion $Ur(t_{sp})$ of the resources is released into the synaptic cleft, and the remaining fraction ($1-r(t)$) becomes available again at rate $\text{1/\ensuremath{\tau} }$. In figure \ref{figure:synapses} a sketch of the neuron-neuron interaction with synaptic depression is illustrated.

It is straightforward to see that when $\tau$ is small the level of synaptic depression is also low since the variable $r(t)$ quickly recover to its maximum value $r_{max}=1$ from the lower values originated by the release of neurotransmitter due to the arrival of a presynaptic spike. When $\tau$ is enlarged such recovering becomes slow and $r(t)$ takes a long time to recover. The synaptic response which is proportional to $r(t)$ will be more depressed the larger the $\tau$. Therefore, we can use $\tau$ as a parameter to control the level of STD.  In classical neural networks including STD, the variable $r(t)$ modulates the synaptic strength, namely $\omega_{ij},$ between the presynaptic $j$ neuron and the postsynaptic $i$ neuron. The postsynaptic neuron receives an input $h_{i}(t)=\omega_{ij}r(t)s_{j}$, where $s_{j}=1,0$ is the neuron state variable  of the input neuron, and which can be seen as an energy term per neuron \cite{amit_89}.

In our quantum case we cannot define the time of spike arrival $t_{sp}$ so we rely on an approximation. In a steady state condition, we can consider that the presynaptic neuron is firing at a given average frequency $f$. In such situation, the second right-hand term of (\ref{eq:dynamic}), after time averaging in such steady state, can be approximated by $U r_{stat} f$ with $r_{stat}=1/(1+\tau U f),$ since one has $f=\langle \sum_{t_{sp}}\delta(t-t_{sp})\rangle=(1/T)\int_{t_0}^{t_0+T} dt \sum_{t_{sp}}\delta(t-t_{sp}),$ being $T$  the temporal window to compute the time average. On the other hand, the presynaptic firing rate $f$ can be interpreted as a measure of the probability for the presynaptic neuron to be firing, in such a way that if $f$ is low the neuron is hardly firing and if $f$ is large the neuron is continuously firing. Hence, we can consider the population of a qubit $\langle \sigma^+ \sigma^-\rangle$ as an appropriate quantum analog to $f$ since it is near to one (i.e. large),  when the qubit in the Up state and zero (low) when the qubit is in the Down state. Therefore, in order to find a quantum analog to the dynamics (\ref{eq:dynamic}) the term including the Dirac delta function in Eq. (\ref{eq:dynamic}) can be approximated by 
\[
Ur_{stat}\langle \sigma^+ \sigma^-\rangle.
\]
In the steady state condition one has
\[
r_{stat}=\frac{1}{1+\tau\, U \, \langle \sigma^+ \sigma^-\rangle}\, ,
\]
which is identical to the classical expression (see above) replacing $f$ by  $\langle \sigma^+ \sigma^-\rangle.$

\begin{figure}[ht!]
\begin{centering}
\includegraphics[width=8cm]{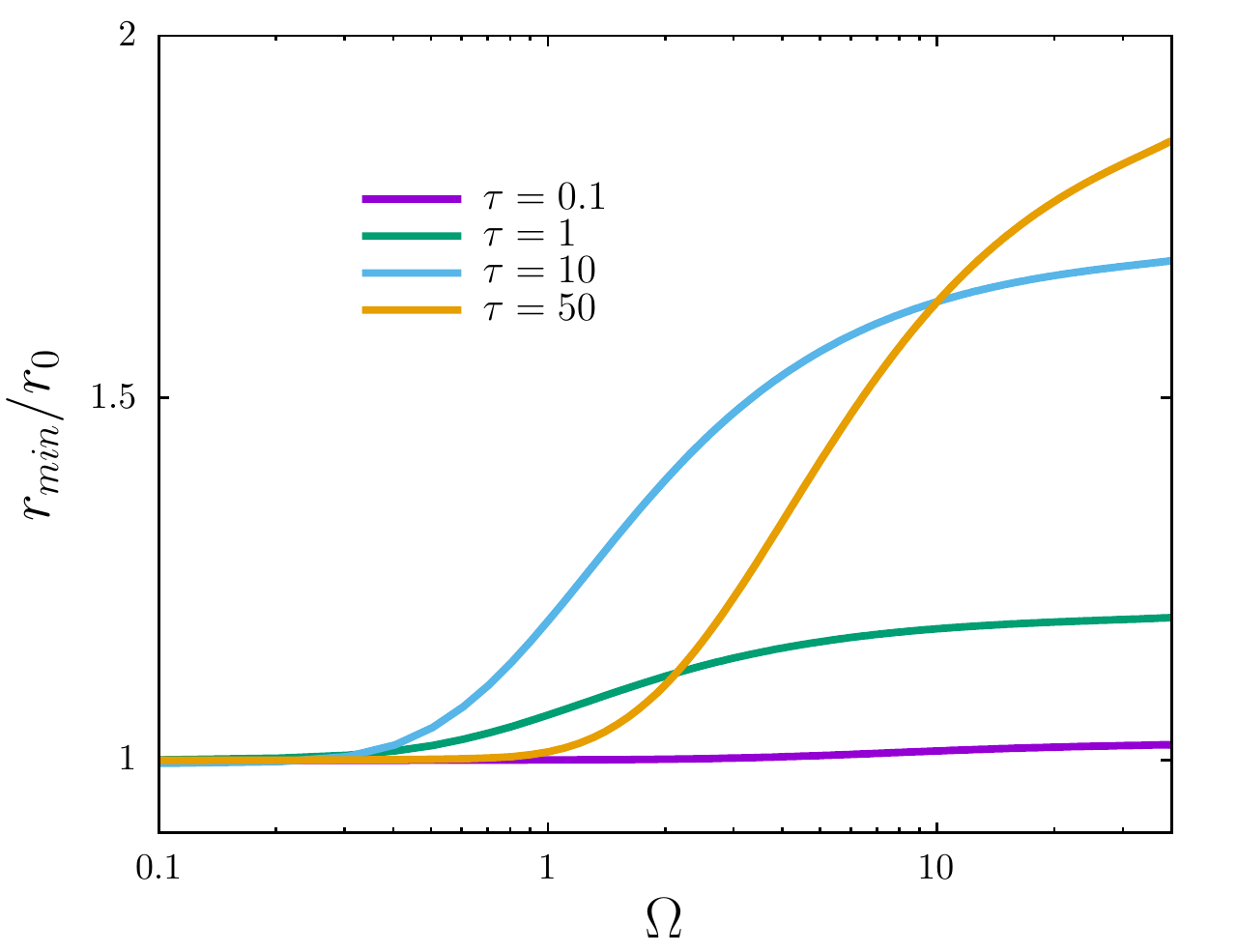}
\par\end{centering}
\caption{The figure illustrates the input-output non-linear features of our system. Renormalized value of the minimum of $r(t)$ reached during its evolution, namely $r_{min},$ as a function of the Hamiltonian parameter $\Omega$ controlling the frequency of the emerging Rabi oscillations of the qubits populations, for different values of the neurotransmitter recovering time $\tau$. Here $r_0=r_{min}(\Omega=0).$ Other Hamiltonian parameters considered are: $U=0.5,\; \epsilon_1=\epsilon_2=0$. This clearly illustrates that the non-linear effects are more important when $\tau$ is large (strong depression) and they disappear for $\tau\rightarrow 0$ (non-depressed case). The value of $r_0=r_{min}(\Omega=0)$ has been calculated after several epochs to avoid transient behaviour. The initial state chosen is $\rho_I=\ket{10}\bra{10}$}
\label{figure:rt}
\end{figure}

Then, for each qubit, one can hypothesise a quantum version of Eq. (\ref{eq:dynamic}) as follow
\begin{equation}
\frac{dr_{i}(t)}{dt}=\frac{1-r_{i}(t)}{\tau}-Ur_{i}(t)  \langle \sigma_i^+ \sigma_i^-\rangle \quad i=1,2\label{eq:plausibleSTD}.
\end{equation}
In  a general  system there are as many values of $r_i$ as neurons. In our two qubit case we will work with only one depression variable.  
We  focus our study to see how the population of qubit $1$ depresses the interaction Hamiltonian. and then substitute in (\ref{eq:ham}) $r=r_{1}(t)$ so the final dynamics of our system is given by a set of Eqs. 
\begin{equation}
\begin{array}{ll}
\frac{d\rho(t)}{dt}= & -i[H(t),\rho(t)],  
\\
\\
\frac{dr(t)}{dt}= & \frac{1-r(t)}{\tau}-Ur(t)\langle \sigma_1^+ \sigma_1^-\rangle(t).
\end{array}\label{eq:final1}
\end{equation}
As it is clear from this set of Eqs. the time-dependent parameter $r(t)$ influences the interaction dynamics of the qubits and it is affected by this dynamics as well. This interaction is non-linear as it is characteristic in classical neuronal systems. To illustrate the emergence of this non-linear effect in our quantum neuron system  we have displayed in Fig. \ref{figure:rt} the minimum value $r_{min}$ that the time-dependent parameter $r(t)$ reaches during its evolution as a function of the parameter $\Omega$ of the Hamiltonian. This is shown for different values of the neurotransmitter recovering time $\tau$. In the figure, such minimum value has been renormalised by dividing  it by the corresponding minimum value when $\Omega=0,$ namely $r_0=r_{min}(\Omega=0), $ in order to make a proper  comparison of the non-linear effects induced by different values of $\tau$. As when $\Omega=0$ the oscillations of the systems are suppressed we select the initial state $\rho_I=\ket{10}\bra{10}$ to ensure a finite population $\left< \sigma^+ \sigma^- \right>=1$. It is clear from the plot that the introduction of the synaptic depression variable $r(t)$ induces a strong non-linear effect in the system for large values of $\tau$. We use the Hamiltonian parameter $\Omega$ as a relevant parameter to tune the input characteristic of our system since it controls the frequency of the Rabi oscillations of the qubits population. 

One difficulty in the design of quantum systems with STD is that the mean value of the excitation of a qubit cannot be estimated by a single measurement, and also the measurement of the population would affect the system state giving rise to a different dynamics. Hence, it is difficult to engineer non-linear dynamics as the one given in Eqs. \ref{eq:final1}. Because of that we have also explored the possibility of implementing STD by a measurement-based protocol that could be easily implemented in realistic systems as cold atoms \cite{jaksch:prl98,diehl:08} and trapped ions \cite{barreiro:nature11,cirac:prl95}.  As the main difficulty of the original model is to design a system with a dynamics that depends on the average population $\left< \sigma^+ \sigma^- \right>$ an easier way to proceed would be to evolve the system according to its Hamiltonian and perform a measurement of the first qubit population. The variable $r(t)$ of the Hamiltonian now follows the dynamics 

\be
\frac{dr(t)}{dt}=\frac{1-r(t)}{\tau}-Ur(t)s_{c},
\ee
where $s_{c}$ is a binary variable which can be $1$ or $0$ depending on the the outcome of the measurement. The algorithm that gives the evolution is now the following:

\begin{enumerate}

\item At the beginning we select $s_c=1$. 

\item For a time $t_m$ the system evolves following the equations 

\begin{equation}
\begin{array}{lll}
\frac{d\rho(t)}{dt} & = & -i[H(t),\rho(t)],  
\\
\\
\frac{dr(t)}{dt}& =&\frac{1-r(t)}{\tau}-Ur(t)s_{c}.
\end{array}\label{eq:algorithm}
\end{equation}

\item  After a time $t_{m}$ the measurement is done, the system collapses and the variable $s_c$ is adjusted to the outcome. From a simulation perspective this is done in the following way: We choose an uniform random number $x_{rand}$ and then make the choice 
\[
s_{c}=\begin{cases}
\begin{array}{c}
1\quad\mbox{if  \ensuremath{x_{rand} < \left< \sigma^+ \sigma^- \right> }}\\
0\quad\mbox{if \ensuremath{x_{rand} >\left< \sigma^+ \sigma^- \right>  }}
\end{array}\end{cases}
\]

\item Go to 2. 

Note that in the simulation we have included a new parameter, $t_m$, that has no correspondence in the continuous evolution. 

\end{enumerate}

\section{Results}

\begin{figure}
\begin{centering}
\includegraphics[width=6cm]{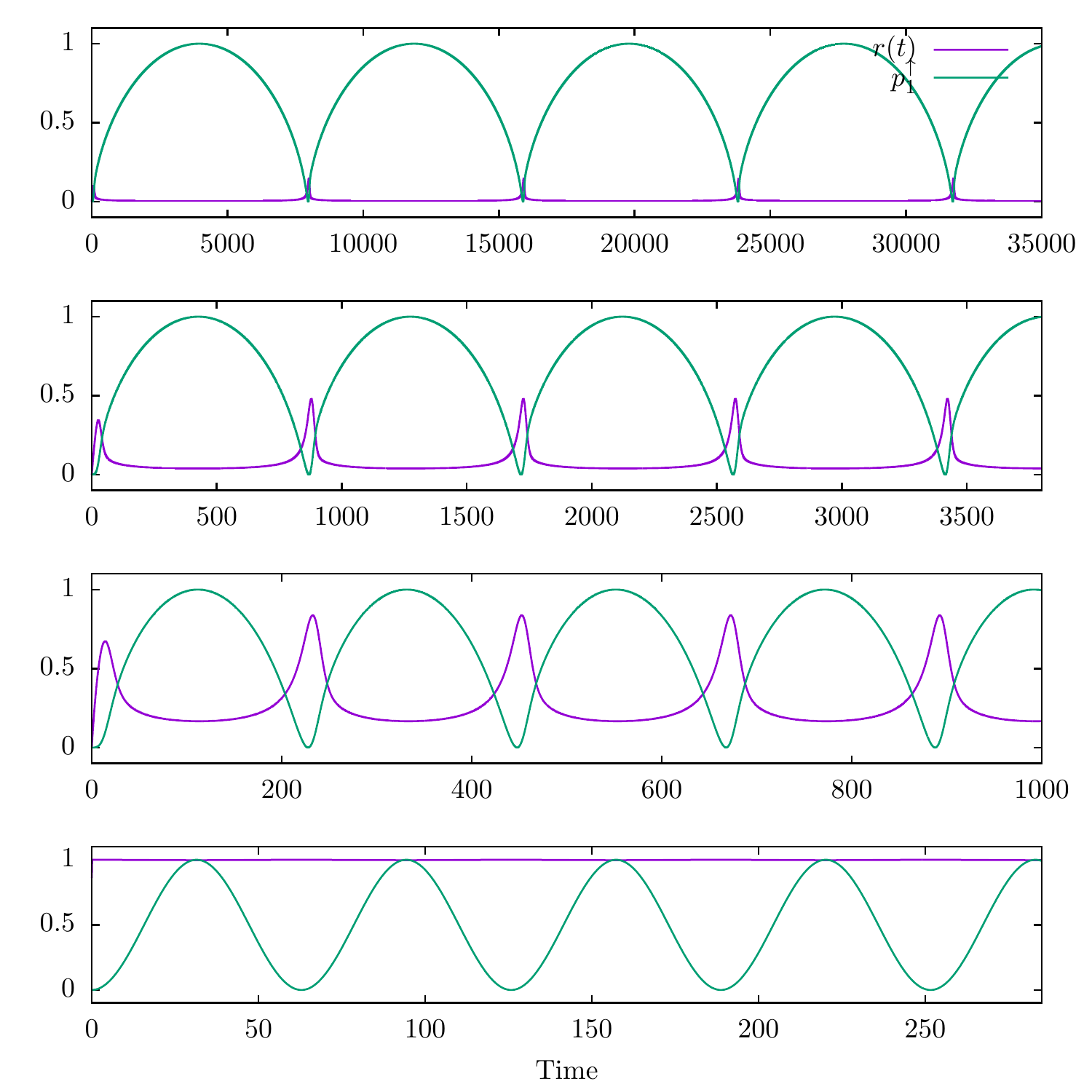}\includegraphics[width=6cm]{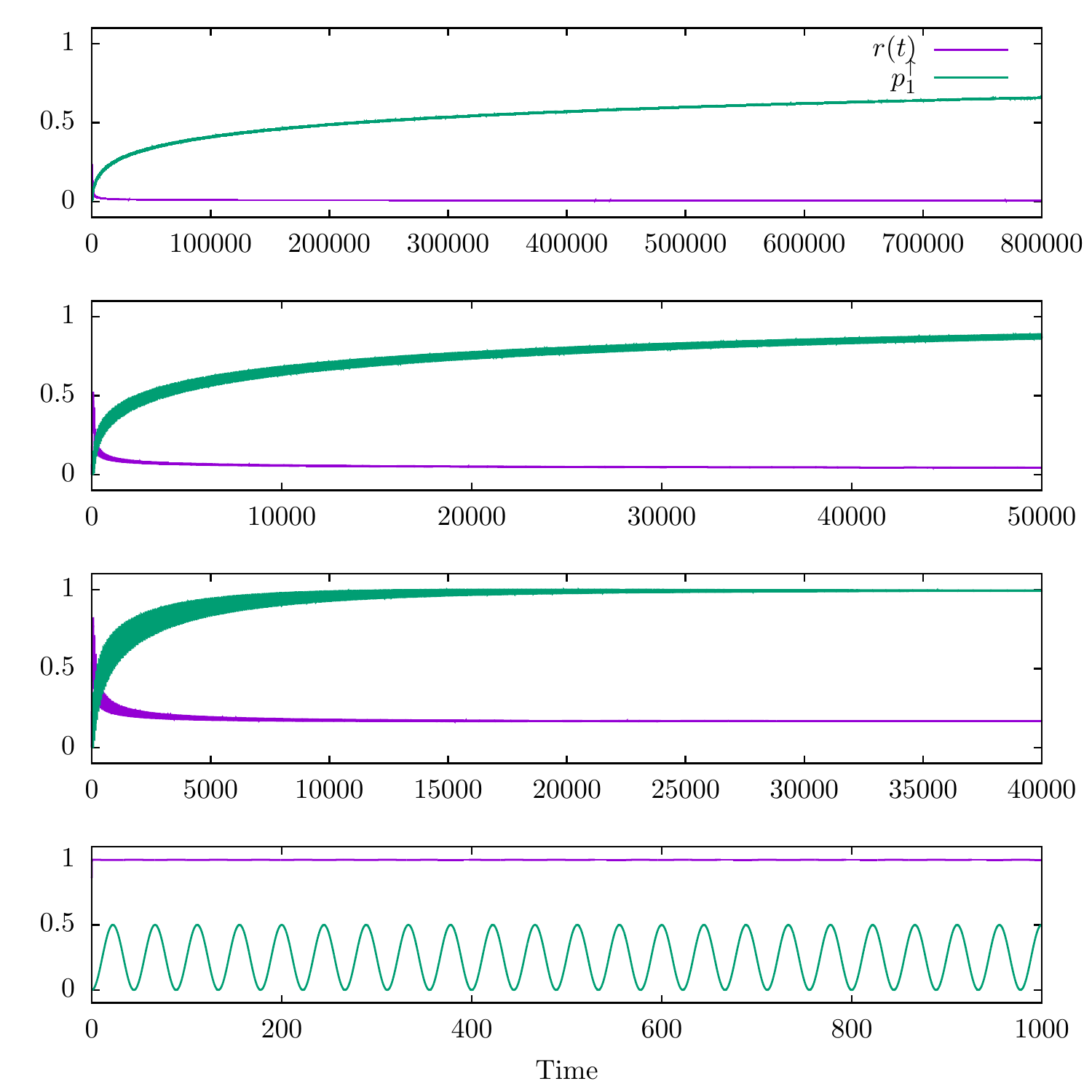}
\par\end{centering}
\caption{Effect of dynamic synapses in an ideal two interacting qubits system. From top to bottom the level of synaptic depression in the system is decreased using, respectively, $\tau=500,\;100,\;10,\;0.001.$ Other parameter values are $U=0.5,\;\Omega=0.05$. In the left panel it is shown the time dependence of the population (green line) of the first qubit as well as the Hamiltonian time-dependent parameter $r(t)$ (purple line) for a symmetric Hamiltonian $\epsilon_{1}=0,\;\epsilon_{2}=0$. In the right panel the same parameters are displayed for $\epsilon_1=0$ and $\epsilon_2=0.1$. The complex interplay among these two variable modulated the shape and frequency of the emergent Rabi oscillations in a nontrivial way. Time given in natural units. Be aware of the different time scales of the plots. The dynamics have been calculated by solving the set of Eqs. (\ref{eq:final1}) with a fourth order Runge Kutta algorithm with time step $\delta t=0.001$ for this and the next plots.  } 
\label{figidealdep}
\end{figure}

We have first analysed how the level of STD in the qubits interaction term affects the behaviour of our 2 qubits system. The results are summarised in Fig. \ref{figidealdep} where the occupation probability of qubit one, $p_{1}^{\uparrow}(t)=\langle \sigma_i^+ \sigma_i^-\rangle(t)$ (green line), is displayed as a function of time as well as $r(t)$ (purple line). We have considered a relatively low interaction strength ($\Omega=0.05$) and an initial state in the form $\rho_{I}=\ket{01}\bra{01}$. In the left panel we observe the case of a balanced system with $\epsilon_1=\epsilon_2=0$. Due to the interchange form of the interaction Hamiltonian both qubits will oscillate between their ground and excited states. In the case of no depression, $\tau\ll 1$, these oscillations are equivalent between both qubits but when the depression time increases a difference between the qubits arises. Qubit one starts being exited for a longer time than qubit two because when it is excited the depression time-dependent parameter $r(t)$ is reduced. This difference increases with $\tau$ and can be used to create a population imbalance between the qubits. Furthermore, if the energy of the qubits is not equal this population imbalance becomes more important, as it is shown in the right panel of Fig. \ref{figidealdep}.  In this case, the qubit responsible of the STD increases its population until reaching a value close to $1$. Interestingly, this behaviour happens for any energy difference, even very small ones. The population imbalance growths faster for medium values of $\tau$. This happens because for small values of $\tau$ the time-dependent parameter $r(t)$ recovers very quickly giving raise to normal Rabi oscillations. On the other hand, when $\tau\gg 1$ the value of $r$ is always close to zero meaning that there is a very weak connection between the qubits, making the growth of the imbalance very slow.  In all cases, once the qubit $1$ is at a population close to one the  variable $r$ drops to zero avoiding the qubit-qubit interaction. Because of that, the system never recovers the Rabi oscillations. We have checked this numerically for times as long as $10^7$ time steps for values of $\tau=10, \, 100,\, 500$. 

\begin{figure}
\begin{centering}
\includegraphics[width=6cm]{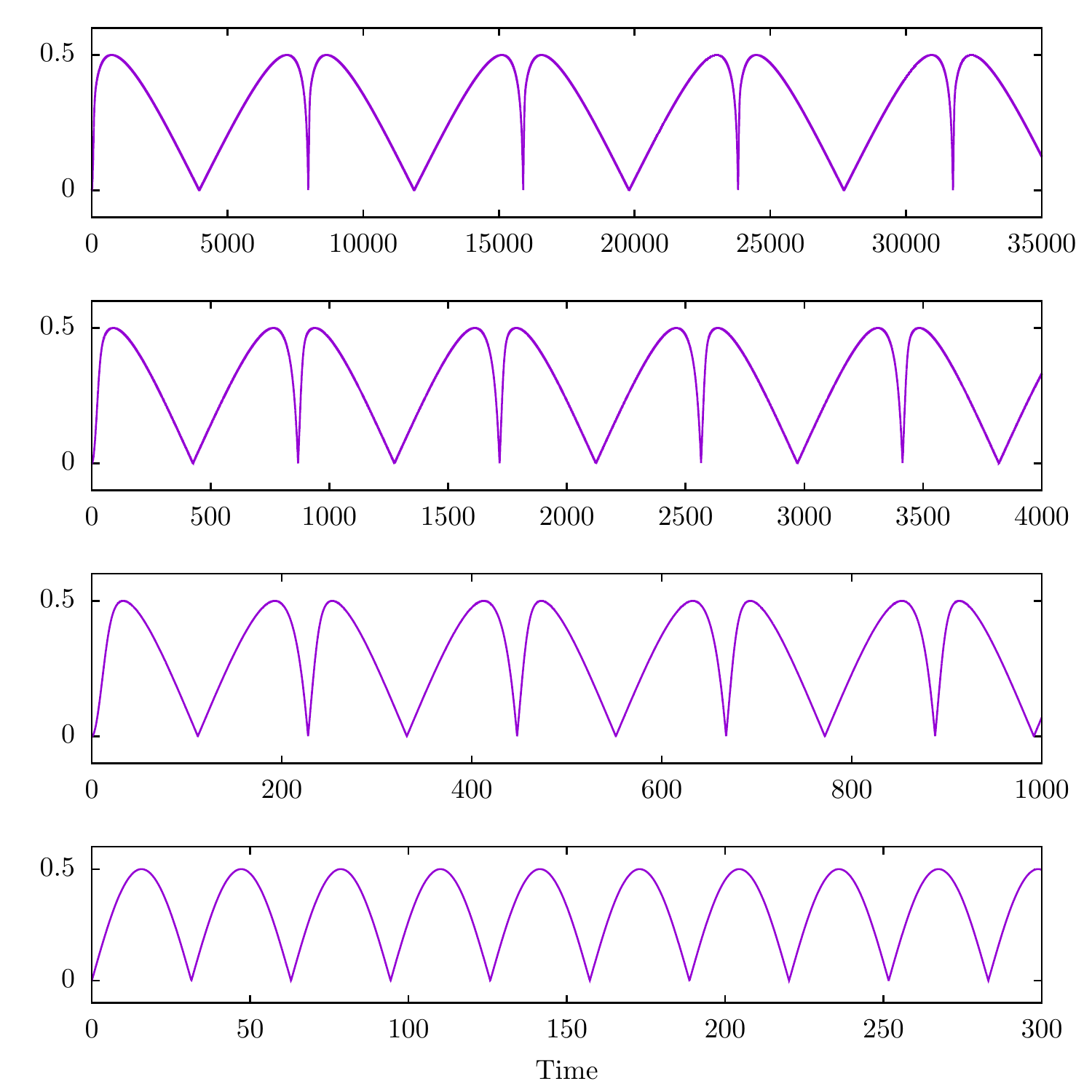}\includegraphics[width=6cm]{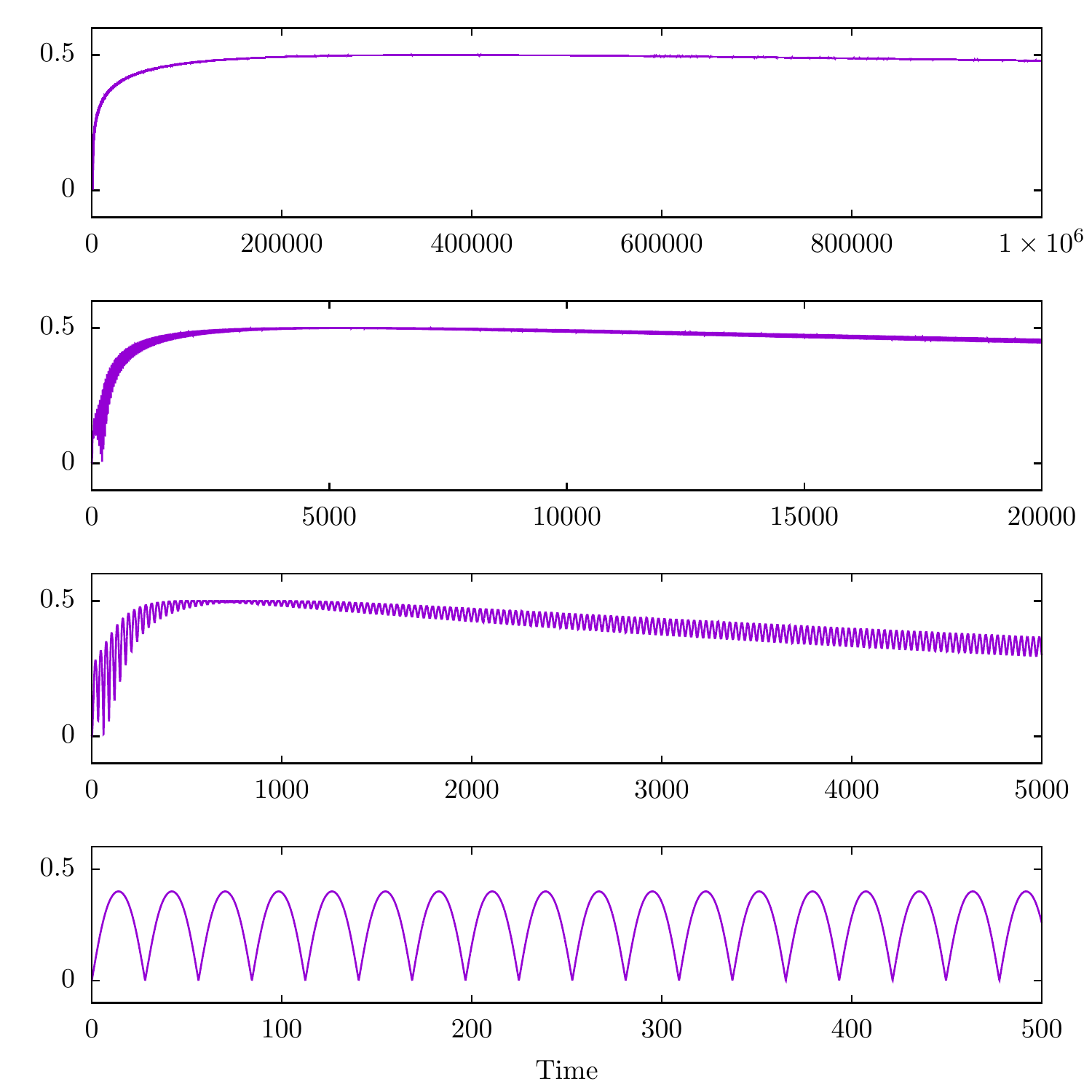}
\par\end{centering}
\caption{Negativity of the system as a function of time. From top to bottom the level of synaptic depression in the system is decreased using, respectively, $\tau=500,\;100,\;10,\;0.001$. In the left panel it is shown for a symmetric Hamiltonian $\epsilon_{1}=0,\;\epsilon_{2}=0$. In the right panel there is an energy imbalance $\epsilon_1=0$ and $\epsilon_2=0.1$. Other parameter values are $U=0.5,\;\Omega=0.05$. Time given in natural units. Be aware of the different time scales of the plots.}
\label{figidealdep2}
\end{figure}

The complex interplay between the qubits dynamics and the depression can be used to create long time entanglement in the system (cf. Fig. \ref{figidealdep2}). In the case with no depression the entanglement between the qubits, measured by the negativity \cite{peres:prl96,horodecki:pla96}, oscillates between $0$ and the maximum value $0.5$ within a time window of $\Delta t\approx 40$. For the energy symmetric case, when the value of the parameter $\tau$ increases the negativity presents a similar behaviour that the first qubit population, increasing the time the system is entangled. This effect, for both the symmetric and asymmetric cases, is displayed in Fig. \ref{figidealdep2}. In the left panel it is shown the balanced case and we can observe that the system still presents oscillations but it is entangled for a longer time than in the non-depression limit (see the plotted time scale in all panels). In the right panel, we observe the case with energy imbalance.  In this case, there is a fast increase of negativity from a separable state to a maximum-entangled one followed by a slow decay of entanglement. The entanglement life is proportional to the STD variable $\tau$, being very long for $\tau\gg 1$.  For all asymmetric cases the entanglement vanishes in the long time limit as in this limit the populations of the qubits reach a steady state. This method to create entanglement using the present biological inspired synaptic mechanism, is deterministic, autonomous, and very long-lived. 

\begin{figure}
\begin{centering}
\includegraphics[width=6cm]{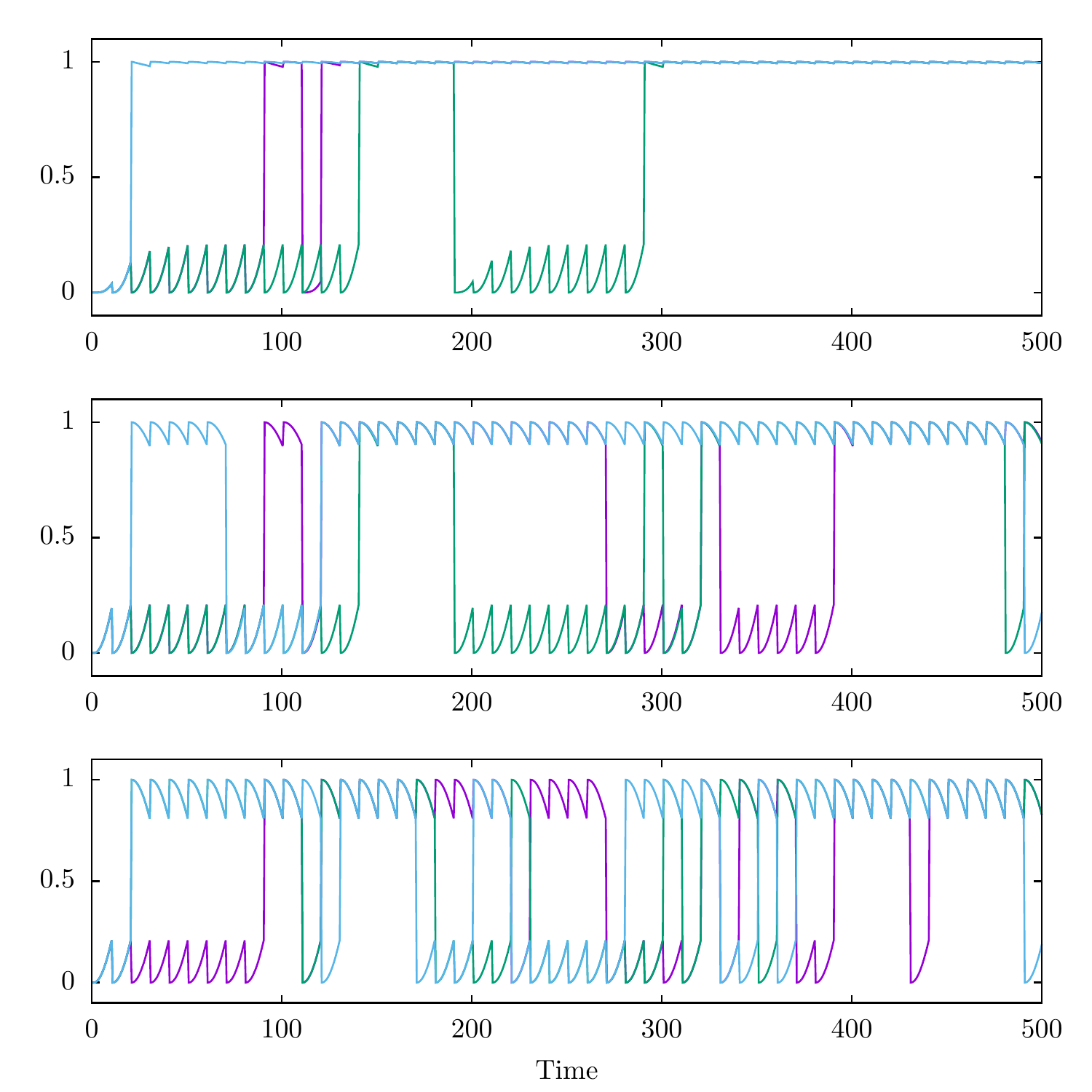}\includegraphics[width=6cm]{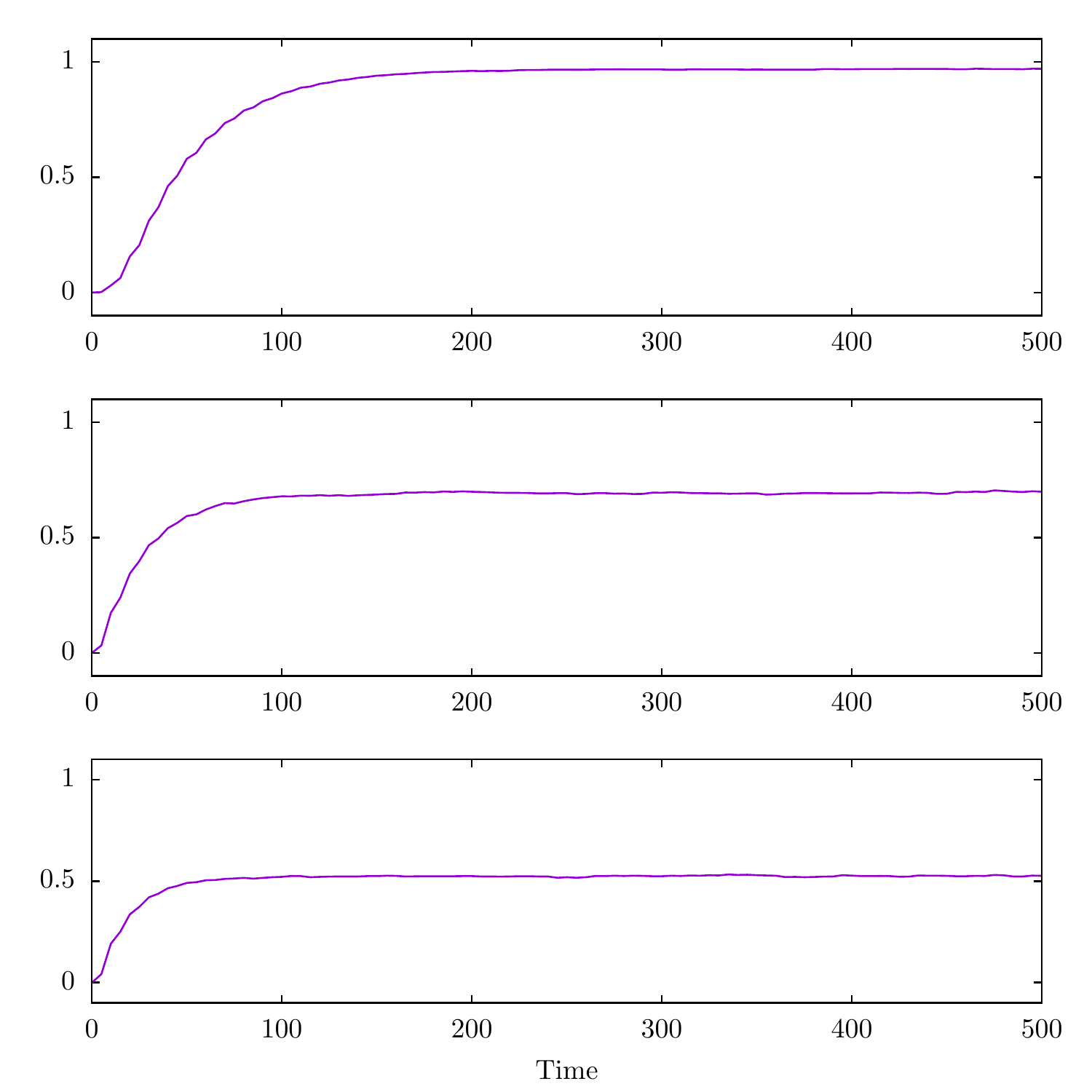}
\par\end{centering}
\caption{Population of the first qubit as a function of time for single trajectories (left) and for the average value over $10000$ trajectories (right). From top to bottom the level of synaptic depression in the system is decreased using, respectively, $\tau=10,\;1,\;0.01$. Other parameter values are $U=0.5,\;\Omega=0.05$,  $\epsilon_1=0$ and $\epsilon_2=0.1$. Time given in natural units and the time between measurements is $t_m=30$ in that units.}
\label{figidealdep3}
\end{figure}

Finally, we have also studied the dynamics of the system under the measurement-based scenario described in the previous Section. In this case, as we cannot know the value of the average population, we perform a measurement of qubit $1$ in the $\left\{\ket{0},\,\ket{1}\right\}$ basis. This measurement makes the system collapse, and depending on its output, we associate to a variable $s_c,$ the value $0$ or $1$. This variable determines the dynamics of the time-dependent depression variable $r_i(t)$. The behaviour of the system is shown in  Fig. \ref{figidealdep3}. In the left panel the population of the first qubit is displayed for three different trajectories for different values of the time-dependent depression parameter $(\tau=10,\; 1, \; 0.01)$, while  the right panel shows the average behaviour. For small values of the depression parameter $\tau$ both qubits perform small oscillations and the measurement process makes stochastic random jumps. In this case the average population of both qubits is $0.5$  as it is expectable because in the limit $\tau\ll 1 $ the system recovers its original quantum Hamiltonian dynamics and the qubits oscilates between the $\ket{0}$ and $\ket{1}$ states. When $\tau$ increases the oscillations when the first qubit is excited are depressed. This makes this configuration more expectable and there are more jumps up than down. The result is a population imbalance that can be appreciated in the average behaviour (right panel). Interestingly, in this measurement-based approach the effect is more relevant for smaller values of $\tau$ than in the previous scenario. This happens because of the binary value of the collapse state variable $s_c$ in this case.

\section{Conclusions}

In this paper we have proposed a novel model of quantum neurons which are interacting via a synaptic depression mechanism. This model is inspired by the synaptic plasticity that happens in biological neuronal networks and it is based in a non-linear interaction of the qubits. This complex interaction allows the emergence of interesting behaviours such as population imbalance between the qubits and long-lasting entanglement generation. These behaviours are more appealing if an energy shift between the qubits is introduced, even if it is small. Furthermore, we have proposed a measurement-based protocol that  can be implemented in real experimental devices. 

This work opens the door to the study of complex quantum neuronal networks with dynamical synapses. Several questions are still open as what is the effect of synapse facilitation, and if quantum dynamics can improve the retrieval capacity of neuronal networks. Furthermore, our two-interacting quantum neurons model can be applied for both feed-forward networks and recurrent networks. It is direct to expand the current study to a perceptron like network since such network is constituted by modules of two-interacting quantum neurons, as the ones considered here, which are integrated in the postsynaptic layer. Moreover, in such a {\em feed forward} perceptron architecture the information flows only in one direction, without interactions between neurons at the same layer. It is natural then to model the synaptic depression as an effect between layers. In this direction, there are some network models  \cite{huber:arxiv_21,pechal:arxiv21,silva:nn20,wiebe:npj16} and the extension of them with more complicated quantum neuron-neuron interaction would be a very interesting field of research. Besides, recurrent networks, as Hopfield's, are also very interesting and have been recently extended to the quantum regime \cite{rotondo:jpa18,rebentrost:pra2018}. Again, the extension of these models to include synaptic depression is straightforward and it would give a more complex, and interesting, behaviour. Finally, the effect of synaptic depression in machine learning problems has already been explored in the classical regime \cite{zhang:ieee16,zhang:sa21}.  The possibility of multipartite entanglement generation by this kind of model is also interesting to study.

\section{Acknowledgements}

This study is part of the Project of I+D+i Ref. PID2020-113681GB-I00, financed by MICIN/AEI/10.13039/501100011033 and FEDER A way to make Europe and also financed by FEDER/Junta de Andaluc\'ia-Consejer\'ia de Transformaci\'on Econ\'omica, Industria, Conocimiento y Universidades/Project Ref. P20.00173. DM wants also to acknowledge funding from the FEDER/Junta de Andaluc\'ia program A.FQM.752.UGR20. 

\bibliography{/Users/daniel_diosdado/Dropbox/investigacion/bibliografia_latex/phys.bib}


\bibliographystyle{unsrt}

\end{document}